\documentclass[12pt]{article}
\usepackage{amsmath}
\usepackage{amssymb}
\usepackage{amsthm}

\theoremstyle{plain}
\newtheorem{thm}{Theorem}
\newtheorem{prop}[thm]{Proposition}

\theoremstyle{definition}
\newtheorem{defn}{Definition}

\theoremstyle{remark}
\newtheorem{rem}{Remark}[section]

\numberwithin{equation}{section}

\title{Two-component soliton systems 
and the Painlev\'e equations}
\author{Mikio Murata\\
Department of Physics and Mathematics, \\
Aoyama Gakuin University, \\
5-10-1 Fuchinobe Sagamihara-shi, \\
Kanagawa 229-8558, Japan
}
\date{}

\begin{document}
\maketitle
\begin{abstract}
We give an extension of the two-component KP hierarchy by 
considering additional time variables. 
We obtain the linear $2\times 2$ system by taking into consideration  
the hierarchy through a reduction procedure. 
The Lax pair of the Schlesinger system and the sixth Painlev\'e 
equation is given from this linear system. 
A unified approach to treat the other Painlev\'e equations from 
the usual two-component KP hierarchy is also considered. 
\end{abstract}



\section{Introduction}
In this paper, we deal with the Painlev\'e equations and 
the soliton systems, and also relations among them. 
Important fact we give attention is correspondence between the 
Painlev\'e equations and the holonomic deformation, 
that is, 
the monodromy preserving deformation of linear differential equations. 
On the other hand, the Kadomtsev-Petviashvili (KP) hierarchy
arises from the isospectral deformation of the eigenvalue problem.
The aim of this paper is to establish correspondence between the 
isospectral deformation and the monodromy preserving deformation. 
We construct an extension of the two-component KP hierarchy by 
introducing new time variables. 
We give also the relation between this hierarchy and the sixth 
Painlev\'e equation. 
We show also the relation between the usual two-component KP hierarchy 
and the other Painlev\'e equations. 

In this introduction, 
we begin by reviewing the theory of the Painlev\'e equations and the 
soliton theory. 
Then we state main results of the present article. 
In Section~\ref{sec:solikp}, 
we construct an extension of the two-component KP hierarchy 
by employing the Sato-Wilson formalism.
In Section~\ref{sec:redkp}, 
we consider the holonomic deformation based on 
this extended hierarchy 
and obtain the nonlinear system that describes the condition of this 
deformation.
We see that the nonlinear system reduces to $\mathrm{P_{VI}}$. 
In Section~\ref{sec:otherkp}, 
we study the holonomic deformation that contains 
the two-component KP hierarchy 
and show that the nonlinear systems that describes the condition of 
this deformation reduce to the other Painlev\'e equations, 
$\mathrm{P_{V}}$, $\mathrm{P_{IV}}$, $\mathrm{P_{III}}$ and 
$\mathrm{P_{II}}$.

\subsection{Painlev\'e equations}
P.~Painlev\'e studied second order ordinary differential equations 
of the form 
\begin{equation}
\frac{d^2y}{dt^2}=F\left(t,y,\frac{dy}{dt}\right)
\end{equation}
where $F$ is analytic in $t$ and rational in $y$ and $dy/dt$.
He tried to determine all of equations without a movable critical 
point. 
P. Painlev\'e and B. Gambier arrived at the following six equations, 
which as known as the Painlev\'e equations (\cite{Gam,P}):
\begin{align}
\mathrm{P_{I}}\colon \frac{d^2y}{dt^2}&=6y^2+t,\\
\mathrm{P_{II}}\colon \frac{d^2y}{dt^2}&=2y^3+ty+\alpha,\\
\mathrm{P_{III}}\colon \frac{d^2y}{dt^2}
&=\frac{1}{y}\left(\frac{dy}{dt}\right)^2
-\frac{1}{t}\frac{dy}{dt}+\frac{\alpha y^2+\beta}{t}
+\gamma y^3+\frac{\delta}{y},\\
\mathrm{P_{IV}}\colon \frac{d^2y}{dt^2}
&=\frac{1}{2y}\left(\frac{dy}{dt}\right)^2
+\frac{3}{2}y^3+4ty^2+2\left(t^2-\alpha\right)y+\frac{\beta}{y},\\
\begin{split}
\mathrm{P_{V}}\colon \frac{d^2y}{dt^2}
&=\left(\frac{1}{2y}+\frac{1}{y-1}\right)
\left(\frac{dy}{dt}\right)^2
-\frac{1}{t}\frac{dy}{dt}
+\frac{(y-1)^2}{t}\left(\alpha y+\frac{\beta}{y}\right)\\
&\quad{}+\frac{\gamma y}{t}+\frac{\delta y(y+1)}{y-1},
\end{split}\\
\begin{split}
\mathrm{P_{VI}}\colon \frac{d^2y}{dt^2}&=
\frac{1}{2}\left(\frac{1}{y}+\frac{1}{y-1}+\frac{1}{y-t}\right)
\left(\frac{dy}{dt}\right)^2-\left(\frac{1}{t}+\frac{1}{t-1}
+\frac{1}{y-t}\right)\frac{dy}{dt}\\
&\quad{}+\frac{y(y-1)(y-t)}{t^2(t-1)^2}
\left[\alpha+\frac{\beta t}{y^2}+\frac{\gamma(t-1)}{(y-1)^2}
+\frac{\delta t(t-1)}{(y-t)^2}\right].
\end{split}
\end{align}
The Painlev\'e equations also appear in the problem of the monodromy 
preserving deformation of linear differential equations. 
R.~Fuchs (\cite{F}) considered the second order linear differential 
equation of Fuchsian type: 
\begin{equation}
\frac{d^2\psi}{d\lambda^2}=p(\lambda,t)\psi
\end{equation}
with the four regular singular points, $\lambda=0,\,1,\,\infty,\,t$ and 
the apparent singularity $\lambda=y$. 
He proved that the sixth Painlev\'e equation, $\mathrm{P_{VI}}$,  
describes the condition that 
the linear differential equation has a fundamental system of solutions 
whose monodromy are independent of a variable $t$. 
Results obtained by R.~Garnier (\cite{G}) is concerned with the 
isomonodromic deformation of the second order linear differential 
equation with irregular singularities. 
He showed that the other five Painlev\'e equations, 
$\mathrm{P_{I}}$, $\mathrm{P_{II}}$, $\mathrm{P_{III}}$, 
$\mathrm{P_{IV}}$, $\mathrm{P_{V}}$, are obtained 
from completely integrability conditions of extended systems of the 
linear differential equation. 
L.~Schlesinger (\cite{S}) considered the isomonodromic deformation of 
the linear system of the first order
differential equations with regular singularities: 
\begin{equation}
\frac{d\Psi}{d \lambda}=\sum_{\nu=1}^n
\frac{A_{\nu}}{\lambda-a_{\nu}}\Psi,\label{eq:sch}
\end{equation}
and obtained the system of nonlinear differential 
equations: 
\begin{align}
\frac{\partial A_\nu}{\partial a_{\mu}}
&=\frac{[A_{\mu},A_{\nu}]}{a_{\mu}-a_{\nu}}\quad(\mu\ne \nu),\\
\frac{\partial A_\nu}{\partial a_{\nu}}
&=-\sum_{\kappa(\ne\nu)}\frac{[A_{\kappa},A_{\nu}]}{a_{\kappa}-a_{\nu}},
\end{align}
where 
\begin{equation*}
[A_{\mu},A_{\nu}]=A_{\mu}A_{\nu}-A_{\nu}A_{\mu}.
\end{equation*}
This system is obtained from the complete integrability condition 
of the extended system of (\ref{eq:sch}):
\begin{align}
\frac{\partial\Psi}{\partial \lambda}&=\sum_{\nu=1}^n
\frac{A_{\nu}}{\lambda-a_{\nu}}\Psi,\\
\frac{\partial\Psi}{\partial a_{\nu}}
&=-\frac{A_{\nu}}{\lambda-a_{\nu}}\Psi.
\end{align}
M.~Jimbo, T.~Miwa and K.~Ueno (\cite{JMU}) established general 
theory of monodromy preserving deformation for the matrix system of 
first order linear ordinary differential equations with regular or 
irregular singularities:  
\begin{equation}
\frac{d\Psi}{d \lambda}
=A(\lambda)\Psi,\label{eq:lini}
\end{equation}
where 
\begin{equation}
A(\lambda)=\sum_{\nu=1}^n\sum_{k=1}^{r_{\nu}}
\frac{A_{\nu,k}}{(\lambda-a_{\nu})^k}
-\sum_{k=2}^{r_{\infty}}A_{\infty,k}\lambda^{k-2}.\label{eq:a}
\end{equation}
In \cite{JMU} monodromy data are being considered as a set of Stokes 
multipliers, 
connection matrices and exponents of formal monodromy, and 
they define the generalized monodromy preserving deformation
is the deformation as monodromy data of a fundamental system 
of solutions are preserved.  
M.~Jim\-bo and T.~Miwa (\cite{JM}) presented the linear systems with 
$2\times 2$ matrices that the Painlev\'e equations are obtained from 
the compatibility condition of them. 
This linear systems are called the Lax pairs for the Painlev\'e 
equations. 
When $A(\lambda)$ (\ref{eq:a}) has a pole of degree $r_{\nu}$ 
at $\lambda=a_{\nu}$, the equation (\ref{eq:lini}) is said to have 
a singular point of Poincar\'e rank $r_{\nu}-1$. 
We associate with each of 
$\lambda=a_{\nu}\ (\nu=1,\dots,n)$ 
a natural number $r_{\nu}$   
such that the Poincar\'e rank of $\lambda=a_{\nu}$ 
is given by $r_{\nu}-1$. 
We also associate with $\lambda=\infty$ 
a natural number $r_{\infty}$  
such that the Poincar\'e rank of $\lambda=\infty$ 
is given by $r_{\infty}-1$. 
Then we can represent such a system of linear differential equations 
by the following symbol: 
\begin{equation}
(r_1,r_2,\dots,r_n,r_{\infty}).
\end{equation}
The system of linear differential equations considered in the studies 
on the Schlesinger system (\ref{eq:sch}) is of the type: 
\begin{equation}
(\underbrace{1,1,\dots,1}_{n+1}).\label{eq:linsch}
\end{equation}
By the use of this notation, the correspondence of the types of 
the linear system with $2\times 2$ 
matrices to the types of the Painlev\'e equations is the following: 
\begin{subequations}
\begin{align}
\mathrm{P_{VI}}&\colon (1,1,1,1),\label{eq:linp6}\\
\mathrm{P_{V}}&\colon (1,1,2),\\
\mathrm{P_{IV}}&\colon (1,3),\\
\mathrm{P_{III}}&\colon (2,2),\\
\mathrm{P_{II}}&\colon (4).
\end{align}
\end{subequations}

\subsection{Soliton systems}
The soliton theory is based on studies of the Korteweg-de 
Vries (KdV) equation.
N.~J.~Zabusky and M.~D.~Kruskal (\cite{ZK}) studied the behavior of 
the numerical solutions of the KdV equation.
They found that the solitary wave solutions had behavior similar 
to the superposition principle, despite the fact that the waves 
themselves were highly nonlinear. 
They named such waves solitons.
This result led C.~S.~Gardner, J.~M.~Greene, M.~D.~Kruskal and 
R.~M.~Miura (\cite{GGKM}) to the discovery of the inverse scattering 
transform method to solve the initial value problems for the KdV 
equation. 
P.~D.~Lax (\cite{L}) showed that the KdV equation is equivalent to 
the isospectral integrability condition for pairs of linear 
operators, known as Lax pairs.
If we introduce the differential operators, 
\begin{align}
L&=\partial_x^2+u,\\
B&=\partial_x^3+\frac{3}{2}u\partial_x+\frac{3}{4}\partial_xu,
\end{align}
then the inverse scattering scheme for the KdV equation is written by 
\begin{align}
L\psi&=\lambda\psi,\label{eq:llin}\\
\partial_t \psi&=B\psi.\label{eq:blin}
\end{align}
If the eigenvalue $\lambda$ is independent of $x$ and $t$, 
then the compatibility condition of the equations (\ref{eq:llin}) 
and (\ref{eq:blin}) yields 
\begin{equation}
\partial_tL=[B,L],
\end{equation}
which reduces to the KdV equation. 
An extension of the Lax equation was given by V.~E.~Zakharov and 
A.~B.~Shabat (\cite{ZS}). 
They treated the following equation for linear differential operators:
\begin{equation}
\partial_yB-\partial_tC+[B,C]=0,\label{eq:zs}
\end{equation}
where
\begin{align}
B&=\sum_{j=0}^{m}b_j\partial_x^j,\\
C&=\sum_{j=0}^{n}c_j\partial_x^j.
\end{align}
The equation (\ref{eq:zs}) is obtained from the compatibility condition of
\begin{align}
\partial_t \psi&=B\psi,\\
\partial_y \psi&=C\psi.\label{eq:clin}
\end{align}
By choosing suitable operators $B$ and $C$, we can obtain 
several soliton equations from the equation (\ref{eq:zs}).
If we put 
\begin{align}
B&=\partial_x^3+\frac{3}{2}u\partial_x+v,\\
C&=\partial_x^2+u, 
\end{align}
then the KP equation is obtained. 
If we suppose that $u$ is independent of $y$, then the KP equation 
reduces to the KdV equation. 
M.~Sato (\cite{Sa,SS}) constructed the KP hierarchy and the 
multi-component KP hierarchy that include the KP equation. 
The solutions of the KP hierarchy constitute an infinite-dimensional 
Grassmann manifold. 
The unified approach to integrability makes us understand 
algebraically and geometrically 
integrable systems with infinitely many degree of freedom and 
their solutions. 
This approach is known as the Sato theory nowadays (\cite{OSTT}). 

\subsection{Relations between soliton systems and Painlev\'e 
equations}
The Painlev\'e equations are treated in the research of 
the mathematical physics. 
It was found by T.~T.~Wu, B.~M.~McCoy, C.~A.~Tracy and E.~Barouch 
(\cite{WMTB}) that the correlation function for the two-dimensional 
Ising model in the scaling region satisfies $\mathrm{P_{III}}$.
In the soliton theory, 
it was demonstrated by M.~J.~Ablowitz and H.~Segur (\cite{AS}) that 
similarity solutions of the soliton equations satisfy the Painlev\'e 
equations. 
The relation between the isomonodromic deformation and 
the isospectral one was discussed; see \cite{FN,JM2,U2,U3}.
M.~Jimbo and T.~Miwa (\cite{JM2}) described a procedure to reduce the 
isospectral deformation into the isomonodromic deformation 
consistently by using the $\tau$-function. 
One can obtain not only the Painlev\'e 
equations themselves but also the Lax pairs of them. 
$\mathrm{P_{III}}$ and $\mathrm{P_{IV}}$ were obtained through 
the reduction from the Pohlmeyer-Lund-Regge equation and 
the nonlinear Schr\"odinger equation, respectively. 
M.~Noumi and Y.~Yamada (\cite{NY}) introduced a Painlev\'e system 
associated with the affine root system of type $A_{n-1}^{(1)}$ 
including $\mathrm{P_{II}}\, (A_{1}^{(1)})$, 
$\mathrm{P_{IV}}\, (A_{2}^{(1)})$ and $\mathrm{P_{V}}\, (A_{3}^{(1)})$. 
The systems are equivalent to similarity reductions of 
the $n$-reduced modified KP hierarchy. 
The coefficients of the Lax pair for the system of type 
$A_{n-1}^{(1)}$ are $n \times n$ matrices (\cite{N}). 
The similarity reductions of the 
Drinfel'd-Sokolov hierarchies was investigated by T.~Ikeda, 
S.~Kakei and T.~Kikuchi; see \cite{KK1,KK2,KK3,KIK}. 
As consequence, 
$\mathrm{P_{V}}$ can be obtained from the modified Yajima-Oikawa 
equation, and 
$\mathrm{P_{VI}}$ with four parameters 
can be derived from the three-wave resonant system.
In the papers, \cite{KK3,KIK}, 
the coefficients of the Lax pair of which they obtained were 
also $3\times 3$ matrices.
They showed that the 
$2\times 2$ linear system can be obtained from 
the $3\times 3$ linear system 
by the method of using the Laplace transformation (\cite{H,M}).

\subsection{Results}
We give systems of the isospectral deformations that 
are directly reduced to the Lax pairs for the Painlev\'e equations. 
Specially, we deal with the linear systems with $2\times 2$ matrices, 
in fact the types of singular points of the linear system with 
$2\times 2$ matrices correspond to the types of the Painlev\'e 
equations.  
Besides reductions of the anti-self-dual Yang-Mills equations to 
ordinary differential equations yield
the Painlev\'e equations. 
The $2\times 2$ linear system of the anti-self-dual Yang-Mills 
equations is also reduced to the Lax pairs for the Painlev\'e 
equations (\cite{MW}).
We intend to study the Painlev\'e equations 
by relating the properties of the soliton equations 
to that of the Painlev\'e equations. 
In order to construct the signpost of this approach, 
we try to formulate the holonomic deformation by using the Sato 
theory. 

In this paper, 
we consider an infinite-dimensional integrable hierarchy and give the 
Lax pair with $2\times 2$ matrices for $\mathrm{P_{VI}}$. 
This hierarchy is an extension of the two-component KP hierarchy 
by using additional time variables.
The extension means that the hierarchy restricted to be independent 
of the introduced time variables is equal to the usual two-component 
KP hierarchy. 
We consider specially the $(1,1)$-reduction of the two-component KP 
hierarchy which is known as the nonlinear Schr\"odinger hierarchy. 
It is contained in the extended Zakharov-Shabat hierarchy; 
cf \cite{D}. 
We formulate the extended hierarchy by using 
the Sato-Wilson formalism and then 
define a wave function  a normal solution of the linear 
system. 
This wave function is of the form similar to the integrand of 
the Lauricella's hypergeometric integral. 

Then we consider the holonomic deformation
in the same way to the isospectral deformation. 
We construct a system of linear differential equations in 
the spectral parameter $\lambda$ by using the wave function 
in the extended hierarchy. 
This linear system is of the type: 
\begin{equation}
(1,1,\dots,1,\infty).\label{eq:linp6g}
\end{equation}
Here, at the infinity, $\lambda=\infty$, the Poincar\'e rank is 
considered as $\infty$ since the coefficient matrix $A(\lambda)$ 
(\ref{eq:a}) contains the formal Laurent series around the point 
$\lambda=\infty$. 
We obtain nonlinear systems that describe the condition of the 
complete integrability of the linear systems. 
If we reduce the type of the linear system (\ref{eq:linp6g}) 
to the type (\ref{eq:linsch}), 
then the infinite-dimensional system is reduced to 
the Schlesinger system, from which $\mathrm{P_{VI}}$ is obtained.

We treat also the other Painlev\'e equations from the viewpoint of 
the usual two-component KP hierarchy. 
We study the nonlinear Schr\"odinger hierarchy 
by using the Sato-Wilson formalism,  
and then give different wave functions, similar to the integrand of 
the some degenerated hypergeometric integral. 
The choice of the wave function can be done freely from the 
two-component KP hierarchy, 
the holonomic deformations might be dependent on it. 
We construct systems of linear differential equations in 
the spectral parameter $\lambda$ by using each wave function. 
The linear systems thus obtained are of the types: 
\begin{subequations}\label{eqs:ling}
\begin{align}
&(1,1,\infty),\\
&(1,\infty),\\
&(2,\infty),\\
&(\infty).
\end{align}
\end{subequations}
We then obtain nonlinear systems that describe the condition of the 
complete integrability of the linear systems. 
If we assume the following reductions for the linear systems 
(\ref{eqs:ling}): 
\begin{subequations}
\begin{align}
(1,1,\infty)&\to(1,1,2),\\
(1,\infty)&\to(1,3),\\
(2,\infty)&\to(2,2),\\
(\infty)&\to(4),
\end{align}
\end{subequations}
then the infinite-dimensional systems is reduced to 
one-dimensional systems which yield the other Painlev\'e equations; 
see Section \ref{sec:otherkp} below. 
It follows that the reductions of the nonlinear Schr\"odinger 
equation give rise to not only $\mathrm{P_{IV}}$ (see \cite{JM2}), 
but also $\mathrm{P_{V}}$ and $\mathrm{P_{III}}$.

\subsection{Remarks}
To study the Painlev\'e equations and the holonomic deformations, 
it is indispensable to consider the extension of the two-component 
KP hierarchy, 
which relates directly to $\mathrm{P_{VI}}$.
F.~Nijhoff, A.~Hone and N.~Joshi have showed that similarity 
reductions of a partial differential equation of Schwarzian type 
(SPDE) lead to $\mathrm{P_{VI}}$ (\cite{NHJ}). 
They gave a Lax pair of $2\times2$ matrices type for the SPDE. 
Therefore it is quite natural to ask that there is an intimate 
relation between the similarity reductions of the SPDE and our 
result. 
This kind of problems remains still open. 
We can also obtain any deformation of a linear differential 
equation with rational coefficients by means of our method.
Studies on such procedure are a main subject of a forthcoming paper. 

\section{An extension of the two-component KP hierarchy}
\label{sec:solikp}
In the present section, 
we study an extension of the $(1,1)$-reduction of 
the two-component KP hierarchy. 
We give a formulation of this hierarchy by using the Sato-Wilson 
formalism, 
and then obtain an integrable system by means of the Zakharov-Shabat 
system. 

\subsection{Pseudo-differential operator}
The multi-component theory of the KP hierarchy is established 
in the paper, \cite{Sa}. 
The $n$-component KP hierarchy is formulated by 
matrix pseudo-differential operators of size $n\times n$, 
instead of scalar ones used in the one-component hierarchy. 
We explain some notation about the matrix pseudo-differential 
operators of size $n\times n$. 

The action of the differential operator $\partial_x$ 
on an $n\times n$ matrix $f(x)$ is 
\begin{equation*}
\partial_xf(x)=\frac{d}{dx}f(x).
\end{equation*}
The operator $\partial_x^{-1}$ is defined by 
\begin{equation*}
\partial_x\partial_x^{-1}=\partial_x^{-1}\partial_x\equiv 1. 
\end{equation*}
Pseudo-differential operators are defined by using the operators 
$\partial_x$ and $\partial_x^{-1}$. 
\begin{defn}
A pseudo-differential operator with matrix-coefficients of 
size $n\times n$ is a linear operator, 
\begin{equation*}
\mathcal{A}=\sum_{m}a_m(x)\partial_x^m, 
\end{equation*}
where $a_m(x)$ is an $n\times n$ matrix-valued function of $x$. 
\end{defn}
A sum of pseudo-differential operators is defined in the usual way 
by collecting terms, 
and their product is defined by the following extension of 
Leibniz's rule, 
\begin{equation*}
\mathcal{A}\mathcal{B}
=\sum_{m,n}a_m(x)\partial_x^m b_n(x)\partial_x^n
=\sum_{m,n}\sum_{k=0}^{\infty}\binom{i}{k}a_m(x)
 b_n^{(m)}(x)\partial_x^{m+n-k},
\end{equation*}
where
\begin{equation*}
\binom{i}{k}=
\begin{cases}
\frac{i(i-1)\dots(i-k+1)}{k!}&(k\ge 1)\\
1&(k=0).
\end{cases}
\end{equation*}
We define the differential operator part of 
a pseudo-differential operator 
$\mathcal{A}$ by
\begin{equation*}
(\mathcal{A})_+=\sum_{m\ge 0}a_m(x)\partial_x^m .
\end{equation*}
A pseudo-differential operator possesses a unique inverse, 
denoted by $\mathcal{A}^{-1}$. 

\subsection{Sato Equation}
In the Sato-Wilson formalism, 
a pseudo-differential operator called the gauge operator
plays an essential role. 
The coefficients of the gauge operator are 
dependent variables in the soliton system. 
The condition of the isospectral deformation is 
given by the Sato equations that the gauge operator should 
satisfy.

We define the gauge operator of size $2\times 2$ by 
\begin{equation}
\mathcal{W}=I+\sum_{k=1}^{\infty}w_{k}\partial_x^{-k},
\end{equation}
whose $2\times 2$ coefficients matrices $w_{k}\,(k\ge 1)$ 
do not depend on the parameter $x$.
This condition for the coefficients is equivalent to 
\lq\lq the $(1,1)$-reduction". 
The formal series $\mathcal{W}$ can be inverted. 
Let 
\begin{equation}
\mathcal{W}^{-1}=\sum_{k=0}^{\infty}v_k\partial_x^{-k},
\end{equation}
the first few $v_k$'s are
\begin{subequations}
\begin{align}
v_0&=I,\\
v_1&=-w_1,\\
v_2&=-w_2+{w_1}^2,\\
v_3&=-w_3+w_1w_2+w_2w_1-{w_1}^3.
\end{align}
\end{subequations}
The gauge operator $\mathcal{W}$ can be used to define the operator
\begin{align}
\begin{split}
\mathcal{U}&=\mathcal{W}\sigma_3\mathcal{W}^{-1}
=\sigma_3+\sum_{k=1}^{\infty}u_k\partial_x^{-k},\label{eq:mkp}
\end{split}
\end{align}
where
\begin{equation*}
\sigma_3=
\begin{pmatrix}
1&0\\
0&-1
\end{pmatrix}
\end{equation*}
and 
\begin{equation}
u_k=\sum_{j=1}^k[w_j,\sigma_3]v_{k-j}\quad(k\ge 1).
\end{equation}
We introduce a differential operator 
\begin{equation}
\mathcal{S}_n=(\gamma_nI+c_n\sigma_3)
\sum_{k=0}^{\infty}{a_n}^{-k-1} \partial_x^k\quad (n=1,\dots,l).
\end{equation}
By employing the gauge operator $\mathcal{W}$ 
and the differential operator $\mathcal{S}_n$,
we define differential operators $\mathcal{B}_n\,(n\ge 1)$ and 
$\mathcal{C}_n\,(n=1,\dots,l)$ by 
\label{eq:bkp}
\begin{align}
\mathcal{B}_n&=\left(\mathcal{W}\sigma_3\partial_x^n
\mathcal{W}^{-1}\right)_+
=\sum_{k=0}^{n-1}u_{n-k}\partial_x^{k}+\sigma_3\partial_x^{n}
\quad (n\ge 1),\\
\mathcal{C}_n&=
\left(\mathcal{W}\mathcal{S}_n\mathcal{W}^{-1}\right)_+
=R_n\sum_{k=0}^{\infty}{a_n}^{-k-1}\partial_x^k \quad (n=1,\dots,l),
\end{align}
where
\begin{equation}\label{subeqs:rkp}
R_n=\gamma_n I+c_n\left(\sigma_3+\sum_{l=1}^{\infty}{a_n}^{-l}u_l\right) 
\quad (n=1,\dots,l).
\end{equation}
Matrix operators
\begin{align}
W&=I+\sum_{k=1}^{\infty}w_{k}\lambda^{-k},\\
U&=\sigma_3+\sum_{k=1}^{\infty}u_k\lambda^{-k},\label{eq:umatkp}\\
S_n&=(\gamma_n I+c_n\sigma_3)\sum_{k=0}^{\infty}{a_n}^{-k-1} \lambda^k
=-\frac{\gamma_nI+c_n\sigma_3}{\lambda-a_n}\quad (n=1,\dots,l),\\
B_n&=\sum_{k=0}^{n-1}u_{n-k}\lambda^k
+\sigma_3\lambda^{n}\quad(n\ge 1),\label{eq:bnmatkp}\\
C_n&=R_n
\sum_{k=0}^{\infty}{a_n}^{-k-1}\lambda^k=-\frac{R_n}{\lambda-a_n}
\quad (n=1,\dots,l)
\label{eq:bmatkp}
\end{align}
are obtained from the pseudo-differential operators by replacing 
$\partial_x$ with $\lambda$.
We assume that the matrix operators satisfy
\begin{align}
\partial_{t_n} W&=B_nW-W\sigma_3\lambda^n\quad(n\ge 1),
\label{eq:satockp} \\
\partial_{a_n} W&=C_nW-WS_n\quad (n=1,\dots,l),\label{eq:satobkp}
\end{align}
which we call the Sato equation hereafter.

Let us now define a wave function.
\begin{defn}
A wave function 
$\Psi(\lambda)$ is defined by the following expression: 
\begin{equation}
\Psi(\lambda)=W\Psi_0(\lambda),\label{eq:wavekp}
\end{equation}
where
\begin{equation}
\begin{split}
\Psi_0(\lambda)
&=\lambda^{\alpha}(\lambda-1)^{\beta}\prod_{n=1}^l(\lambda-a_n)^{\gamma_n}
\exp(x\lambda)\\
&\quad{}\times\mathop{\mathrm{diag}}\Biggl\{
\lambda^{a}(\lambda-1)^{b}\prod_{n=1}^l(\lambda-a_n)^{c_n}
\exp\left(\sum_{n=1}^{\infty}t_n\lambda^n\right),\\
&\qquad\quad{}\lambda^{-a}(\lambda-1)^{-b}\prod_{n=1}^l(\lambda-a_n)^{-c_n}
\exp\left(-\sum_{n=1}^{\infty}t_n\lambda^n\right)\Biggr\}.
\end{split}\label{eq:psi0kp}
\end{equation}
\end{defn}
The elements of the wave function are similar to 
the integrand of the Lauricella's hypergeometric integral:
\begin{equation}
\begin{split}
&F_D(a,b_1,\dots,b_l,c;a_1,\dots,a_l)\\
&=\frac{\Gamma(c)}{\Gamma(a)\Gamma(c-a)}
\int_0^1\lambda^{a-1}(1-\lambda)^{c-a-1}
\prod_{n=1}^l(1-a_n\lambda)^{-b_n}d\lambda.
\end{split}
\end{equation}
We note that the matrix-valued function $\Psi_0(\lambda)$ satisfies 
\begin{align}
\partial_x\Psi_0(\lambda)&=\lambda\Psi_0(\lambda),\\
\partial_{t_n}\Psi_0(\lambda)
&=\sigma_3 \lambda^n\Psi_0(\lambda)
=\sigma_3\partial_x^n\Psi_0(\lambda)\quad(n\ge 1),\\
\partial_{a_n}\Psi_0(\lambda)&=S_n\Psi_0(\lambda)
=\mathcal{S}_n\Psi_0(\lambda)\quad (n=1,\dots,l).
\end{align}

This leads to the following theorem: 
\begin{thm}\label{prop:laxkp}
If a matrix operator $W$ satisfies 
the Sato equation $(\ref{eq:satockp})$ and $(\ref{eq:satobkp})$,
then the wave function $\Psi(\lambda)$ 
which can be derived from $W$ satisfies the linear systems, 
\begin{align}
\partial_x\Psi(\lambda)
&=\lambda\Psi(\lambda),\label{eq:linlkp} \\ 
\partial_{t_n}\Psi(\lambda)
&=B_n\Psi(\lambda)\quad(n\ge 1),\label{eq:linckp} \\ 
\partial_{a_n} \Psi(\lambda)
&=C_n\Psi(\lambda)\quad (n=1,\dots,l).\label{eq:linbkp}
\end{align}
\end{thm}

\begin{proof} 
We have
\begin{equation}
\begin{split}
\partial_x\Psi(\lambda)-\lambda\Psi(\lambda)
&=\partial_x\left(W\Psi_0(\lambda)\right)-\lambda W\Psi_0(\lambda)\\
&=(\partial_xW)\Psi_0(\lambda)\\
&=0,
\end{split}
\end{equation}
since $\partial_xW=0$. We find
\begin{equation}
\begin{split}
\partial_{t_n}\Psi(\lambda)-B_n\Psi(\lambda)
&=\partial_{t_n}\left(W\Psi_0(\lambda)\right)-B_nW\Psi_0(\lambda)\\
&=\left(\partial_{t_n}W-B_nW+W\sigma_3\lambda^{n}\right)\Psi_0(\lambda)\\
&=0
\end{split}
\end{equation}
by the Sato equation (\ref{eq:satockp}). We obtain
\begin{equation}
\begin{split}
\partial_{a_n}\Psi(\lambda)-C_n\Psi(\lambda)
&=\partial_{a_n}\left(W\Psi_0(\lambda)\right)-C_nW\Psi_0(\lambda)\\
&=\left(\partial_{a_n}W-C_nW+WS_n\right)\Psi_0(\lambda)\\
&=0
\end{split}
\end{equation}
by the Sato equation (\ref{eq:satobkp}). 
\end{proof}

The Sato equations also lead to the following theorem: 
\begin{thm}\label{prop:zskp}
If a matrix operator $W$ satisfies 
the Sato equation $(\ref{eq:satockp})$ and $(\ref{eq:satobkp})$, 
then the matrix operators $U$, $B_n$ and $C_n$
satisfy the Lax-type systems, 
\begin{align}
\partial_{t_n} U&=[B_n,U]\quad (n\ge 1),\\
\partial_{a_n} U&=[C_n,U]\quad (n=1,\dots,l),
\end{align}
and the Zakharov-Shabat systems,
\begin{align}
\partial_{t_m}B_n-\partial_{t_n}B_m+[B_n,B_m]&=0\quad (n,m\ge 1),
\label{eq:zs2kp}\\
\partial_{a_m}B_n-\partial_{t_n}C_m+[B_n,C_m]&=0\quad (n\ge 1,\,m=1,\dots,l),
\label{eq:zskp}\\
\partial_{a_m}C_n-\partial_{a_n}C_m+[C_n,C_m]&=0\quad (n,m=1,\dots,l).
\label{eq:zs3kp}
\end{align}
\end{thm}

\begin{proof}
From the definition of the pseudo-differential operator 
$\mathcal{U}$ (\ref{eq:mkp}), we find 
\begin{equation}
U=W\sigma_3W^{-1}.
\end{equation}
Therefore we have 
\begin{equation}
\begin{split}
\partial_{t_n} U-[B_n,U]
&=\partial_{t_n} (W\sigma_3W^{-1})-[B_n,W\sigma_3W^{-1}]\\
&=\left[\left(\partial_{t_n} W-B_nW+W\sigma_3\lambda^n\right)
W^{-1},W\sigma_3W^{-1}\right]\\
&=0
\end{split}
\end{equation}
by the Sato equation (\ref{eq:satockp}). 
We obtain 
\begin{equation}
\begin{split}
\partial_{a_n} U-[C_n,U]
&=\partial_{a_n} (W\sigma_3W^{-1})-[C_n,W\sigma_3W^{-1}]\\
&=\left[\left(\partial_{a_n} W-C_nW+WS_n\right)W^{-1},W\sigma_3W^{-1}\right]\\
&=0
\end{split}
\end{equation}
by the Sato equation (\ref{eq:satobkp}).
We find
\begin{equation}
\begin{split}
&\partial_{t_m}B_n-\partial_{t_n}B_m+[B_n,B_m]\\
&=-\partial_{t_m}\left\{\left(\partial_{t_n} W-B_nW
+W\sigma_3\lambda^n\right)
W^{-1}\right\}\\
&\quad{}+\partial_{t_n}\left\{\left(\partial_{t_m} W-B_mW
+W\sigma_3\lambda^m\right)W^{-1}\right\}\\
&\quad{}-\left[B_n,\left(\partial_{t_m} W-B_mW+W
\sigma_3\lambda^m\right)W^{-1}\right]\\
&\quad{}-\left[\left(\partial_{t_n} W-B_nW
+W\sigma_3\lambda^n
\right)W^{-1},\left(
\partial_{t_m}W+W\sigma_3\lambda^m\right)W^{-1}\right]\\
&=0
\end{split}
\end{equation}
by the Sato equations (\ref{eq:satockp}). 
We see
\begin{equation}
\begin{split}
\partial_{a_m}B_n&-\partial_{t_n}C_m+[B_n,C_m]\\
&=-\partial_{a_m}\left\{\left(\partial_{t_n} W-B_nW
+W\sigma_3\lambda^n\right)
W^{-1}\right\}\\
&\quad{}+\partial_{t_n}\left\{\left(\partial_{a_m} W-C_mW+WS_m\right)
W^{-1}\right\}\\
&\quad{}-\left[B_n,\left(\partial_{a_m} W-C_mW+WS_m\right)W^{-1}\right]\\
&\quad{}-\left[\left(\partial_{t_n} W-B_nW
+W\sigma_3\lambda^n
\right)W^{-1},
\left(\partial_{a_m}W+WS_m\right)W^{-1}\right]\\
&=0
\end{split}
\end{equation}
by the Sato equations (\ref{eq:satockp}) and (\ref{eq:satobkp}). 
We have 
\begin{equation}
\begin{split}
\partial_{a_m}C_n&-\partial_{a_n}C_m+[C_n,C_m]\\
&=-\partial_{a_m}\left\{\left(\partial_{a_n} W-C_nW
+WS_n\right)
W^{-1}\right\}\\
&\quad{}+\partial_{a_n}\left\{\left(\partial_{a_m} W-C_mW+WS_m\right)
W^{-1}\right\}\\
&\quad{}-\left[C_n,\left(\partial_{a_m} W-C_mW+WS_m\right)W^{-1}\right]\\
&\quad{}-\left[\left(\partial_{a_n} W-C_nW
+WS_n
\right)W^{-1},
\left(\partial_{a_m}W+WS_m\right)W^{-1}\right]\\
&=0
\end{split}
\end{equation}
by the Sato equation (\ref{eq:satobkp}).
\end{proof}

The systems (\ref{eq:zs2kp}) are equal to the 
Zakharov-Shabat systems in 
the $(1,1)$-reduction of the two-component KP hierarchy. 
The systems (\ref{eq:zskp}) and (\ref{eq:zs3kp}) are the additional 
ones in the extended hierarchy. 
So new integrable systems are obtained from the systems 
(\ref{eq:zskp}) and (\ref{eq:zs3kp}).
Since the left-hand side of (\ref{eq:zskp}) is
\begin{equation}
\begin{split}
\partial_{a_m}C_n&-\partial_{a_n}C_m+[C_n,C_m]\\
&=\left(\partial_{a_n}R_m-\frac{[R_n,R_m]}{a_n-a_m}\right)
\frac{1}{\lambda-a_m}
-\left(\partial_{a_m}R_n-\frac{[R_n,R_m]}{a_n-a_m}\right)
\frac{1}{\lambda-a_n},
\end{split}
\end{equation}
we obtain 
\begin{equation}
\partial_{a_n}R_m-\frac{[R_n,R_m]}{a_n-a_m}=0.
\end{equation}
Since the left-hand side of (\ref{eq:zskp}) is
\begin{equation}
\begin{split}
\partial_{a_m}B_n&-\partial_{t_n}C_m+[B_n,C_m]\\
&=\left(\partial_{t_n}R_m-
\left[\sum_{l=0}^{n-1}{a_m}^lu_{n-l}+{a_m}^n\sigma_3,R_m\right]\right)
\frac{1}{\lambda-{a_m}}\\
&{}+\sum_{k=1}^{n}
\left(\partial_{a_m}u_k-\left[\sum_{l=1}^{k-1}{a_m}^{l-1}u_{k-l}
+{a_m}^{k-1}\sigma_3,R_m\right]\right)\lambda^{n-k},
\end{split}
\end{equation}
we obtain systems
\begin{subequations}\label{subeqs:solkp}
\begin{align}
\partial_{t_n}R_m-\left[\sum_{l=1}^{n}{a_m}^{n-l}u_{l}
+{a_m}^n\sigma_3,R_m\right]&=0,\\
\partial_{a_m}u_k-\left[\sum_{l=1}^{k-1}{a_m}^{k-l-1}u_{l}
+{a_m}^{k-1}\sigma_3,R_m\right]&=0\quad(k=1.\dots n).
\end{align}
\end{subequations}
If we set $n=1$, then 
the system (\ref{subeqs:solkp}) reduces to 
\begin{subequations}
\begin{align}
\partial_{t_1}R_m-\left[u_{1}+a_m\sigma_3,R_m\right]&=0,\\
\partial_{a_m}u_1-\left[\sigma_3,R_m\right]&=0.
\end{align}
\end{subequations}
If we introduce the following parameterizations for the matrices 
\begin{subequations}
\begin{align}
u_1&=\begin{pmatrix}
0&u\\
v&0
\end{pmatrix},\label{eq:m1kp}\\
R_m&=\begin{pmatrix}
g+h&e\\
f&g-h
\end{pmatrix},
\end{align}
\end{subequations}
then we obtain a system
\begin{subequations}
\begin{align}
\partial_{a_m}u-2e&=0,\\
\partial_{a_m}v+2f&=0,\\
\partial_{t_1}e-2a_me+2hu&=0,\\
\partial_{t_1}f+2a_mf-2hv&=0,\\
\partial_{t_1}h-fu+ev&=0,\\
\partial_{t_1}g&=0.
\end{align}
\end{subequations}

\begin{rem}
We have formulated the hierarchy by using the 
pseudo-dif\-fer\-en\-tial operators. 
We can also formulate that by using the difference operators 
(see \cite{UT}). 
If the gauge operator $\mathcal{W}$ does not depend on the 
parameter $\alpha$, then we have
\begin{equation}
e^{\partial_{\alpha}}\Psi(\lambda)=\lambda\Psi(\lambda).
\end{equation}
Therefore the difference operators are obtained from 
the pseudo-differential operators by replacing 
$\partial_x$ with $e^{\partial_{\alpha}}$.
\end{rem}

\section{The extended two-component system and the sixth Painlev\'e 
equation}
\label{sec:redkp}
In this section, 
we consider a holonomic deformation of systems, 
obtained from the integrable system given in the previous section. 
We construct a system of linear differential equations in 
the spectral parameter $\lambda$ by using the wave function 
in the extended hierarchy, 
and then obtain nonlinear systems that describe the condition of the 
complete integrability of the linear systems. 
We show that the infinite-dimensional system is reduced to 
the Schlesinger system, from which $\mathrm{P_{VI}}$ is obtained.

If we introduce a differential operator
\begin{equation}
\begin{split}
\mathcal{V}
&=I\left(\alpha
-\beta\sum_{k=1}^{\infty}\partial_x^k
-\sum_{n=1}^l\gamma_n\sum_{k=1}^{\infty}{a_n}^{-k}\partial_x^k
+x\partial_x\right)\\
&\quad{}+\sigma_3\left(a
-b\sum_{k=1}^{\infty}\partial_x^k
-\sum_{n=1}^lc_n\sum_{k=1}^{\infty}{a_n}^{-k}\partial_x^k
+\sum_{n=1}^{\infty}nt_n\partial_x^{n}\right), 
\end{split}
\end{equation}
then the matrix-valued function $\Psi_0(\lambda)$ 
(\ref{eq:psi0kp}) fulfills 
\begin{equation}
\lambda\partial_{\lambda}\Psi_0(\lambda)
=\mathcal{V}\Psi_0(\lambda). 
\end{equation}
By using the gauge operator $\mathcal{W}$ 
and the differential operator $\mathcal{V}$,
we define a differential operator $\mathcal{D}$ by 
\begin{equation}
\mathcal{D}=
\left(\mathcal{W}\mathcal{V}\mathcal{W}^{-1}\right)_{+}
=\sum_{k=0}^{\infty}d_k\partial_x^{k},
\end{equation}
where
\begin{subequations}
\begin{align}
d_0&=\alpha I+a \sigma_3-b\sum_{l=1}^{\infty}u_l
-\sum_{n=1}^lc_n\sum_{l=1}^{\infty}{a_n}^{-l}u_l
+\sum_{n=1}^{\infty}nt_nu_n,\\
\begin{split}
d_1&=\left(-\beta-\sum_{n=1}^l\gamma_n {a_n}^{-1}+x\right)I
-b\left(\sigma_3+\sum_{l=1}^{\infty}u_l\right)\\
&\quad{}-\sum_{n=1}^lc_n{a_n}^{-1}
\left(\sigma_3+\sum_{l=1}^{\infty}{a_n}^{-l}u_l\right)
+t_1\sigma_3+\sum_{n=2}^{\infty}nt_nu_{n-1},
\end{split}\\
\begin{split}
d_k&=\left(-\beta -\sum_{n=1}^l\gamma_n {a_n}^{-k}\right)I
-b\left(\sigma_3+\sum_{l=1}^{\infty}u_l\right)\\
&\quad{}-\sum_{n=1}^lc_n{a_n}^{-k}
\left(\sigma_3+\sum_{l=1}^{\infty}{a_n}^{-l}u_l\right)
+kt_k\sigma_3+\sum_{n=k+1}^{\infty}nt_nu_{n-k}\quad(k\ge 2).
\end{split}
\end{align}
\end{subequations}
We introduce matrix operators
\begin{gather}
T=\frac{\alpha I+a\sigma_3}{\lambda}
+\frac{\beta I+b\sigma_3}{\lambda-1}
+\sum_{n=1}^l\frac{\gamma_n I+c_n\sigma_3}{\lambda-a_n}
+\sum_{n=1}^{\infty}nt_n\sigma_3\lambda^{n-1},\\ 
A=\sum_{k=0}^{\infty}d_k\lambda^{k-1}.
\end{gather}
We note that 
\begin{equation}
\partial_{\lambda}\Psi_0(\lambda)=T\Psi_0(\lambda). 
\end{equation}
We assume that the matrix operator $A$ satisfies 
the Sato equation with respect to the spectral parameter:
\begin{equation}\label{eq:satoakp}
\partial_\lambda W=AW-WT.
\end{equation}

This leads to the following theorem: 
\begin{thm}\label{prop:laxakp}
If a matrix operator $W$ satisfies the reduction condition 
$(\ref{eq:satoakp})$, 
then the wave function $\Psi(\lambda)$ $(\ref{eq:wavekp})$ 
satisfies the linear system
\begin{equation}\label{eq:linakp}
\partial_\lambda \Psi(\lambda)=A\Psi(\lambda).
\end{equation}
\end{thm}

\begin{proof}
We have 
\begin{equation}
\begin{split}
\partial_\lambda\Psi(\lambda)-A\Psi(\lambda)
&=\partial_\lambda\left(W\Psi_0(\lambda)\right)-AW\Psi_0(\lambda)\\
&=\left(\partial_\lambda W-AW+WT\right)\Psi_0(\lambda)\\
&=0
\end{split}
\end{equation}
by the condition (\ref{eq:satoakp}). 
\end{proof}

The Sato equations also lead to the following theorem: 
\begin{thm}\label{prop:zsakp}
If a matrix operator $W$ satisfies 
the Sato equation $(\ref{eq:satockp})$, $(\ref{eq:satobkp})$ 
and $(\ref{eq:satoakp})$, 
then the matrix operators $U$ and $A$ 
satisfy the Lax-type systems, 
\begin{equation}
\partial_{\lambda} U=[A,U],
\end{equation}
and the matrix operators $A$, $B_n$ and $C_n$ satisfy 
the Zakharov-Shabat type systems,
\begin{align}
\partial_{t_n}A-\partial_{\lambda}B_n+[A,B_n]&=0\quad (n\ge 1),
\label{eq:cc2kp}\\
\partial_{a_n}A-\partial_{\lambda}C_n+[A,C_n]&=0\quad (n=1,\dots,l).
\label{eq:cckp}
\end{align}
\end{thm}

\begin{proof}
We have 
\begin{equation}
\begin{split}
\partial_{\lambda} U-[A,U]
&=\partial_{\lambda} (W\sigma_3W^{-1})-[A,W\sigma_3W^{-1}]\\
&=\left[\left(\partial_{\lambda} W-AW+WT\right)W^{-1},
W\sigma_3W^{-1}\right]\\
&=0
\end{split}
\end{equation}
by the Sato equation (\ref{eq:satoakp}). 
We find
\begin{equation}
\begin{split}
\partial_{t_n}A&-\partial_{\lambda}B_n+[A,B_n]\\
&=-\partial_{t_n}\left\{\left(\partial_{\lambda} W-AW+WT\right)
W^{-1}\right\}\\
&\quad{}+\partial_{\lambda}\left\{\left(\partial_{t_n} W
-B_nW+W\sigma_3\lambda^n\right)W^{-1}\right\}\\
&\quad{}-\left[A,\left(\partial_{t_n} W-B_nW+W
\sigma_3\lambda^n\right)W^{-1}\right]\\
&\quad{}-\left[\left(\partial_{\lambda} W-AW+WT\right)W^{-1},\left(
\partial_{t_n}W+W\sigma_3\lambda^n\right)W^{-1}\right]\\
&=0
\end{split}
\end{equation}
by the Sato equations (\ref{eq:satockp}) and (\ref{eq:satoakp}). 
We have 
\begin{equation}
\begin{split}
\partial_{a_n}A&-\partial_{\lambda}C_n+[A,C_n]\\
&=-\partial_{a_n}\left\{\left(\partial_{\lambda} W-AW+WT\right)
W^{-1}\right\}\\
&\quad{}+\partial_{\lambda}
\left\{\left(\partial_{a_n} W-C_nW+WS_n\right)W^{-1}\right\}\\
&\quad{}-\left[A,\left(\partial_{a_n} W-C_nW+WS_n\right)W^{-1}\right]\\
&\quad{}-\left[\left(\partial_{\lambda} W-AW+WT\right)W^{-1},
\left(\partial_{a_n}W+WS_n\right)W^{-1}\right]\\
&=0
\end{split}
\end{equation}
by the Sato equations (\ref{eq:satobkp}) and (\ref{eq:satoakp}).
\end{proof}

If we introduce matrices 
\begin{subequations}
\begin{align}
P&=\alpha I+a \sigma_3-b\sum_{l=1}^{\infty}u_l
-\sum_{n=1}^lc_n\sum_{l=1}^{\infty}{a_n}^{-l}u_l
+\sum_{n=1}^{\infty}nt_nu_n\\
Q&=\beta I+b\left(\sigma_3+\sum_{l=1}^{\infty}u_l\right),\\
T_0&=xI+t_1\sigma_3+\sum_{n=2}^{\infty}nt_nu_{n-1},\\
T_k&=(k+1)t_{k+1}\sigma_3+\sum_{n=k+2}^{\infty}nt_nu_{n-k-1}\quad(k\ge 1),
\end{align}
\end{subequations}
then we have 
\begin{equation}
\begin{split}
A=\frac{P}{\lambda}+\frac{Q}{\lambda-1}
+\sum_{n=1}^l\frac{R_n}{\lambda-a_n}
+\sum_{k=0}^{\infty}T_k\lambda^k,
\end{split}\label{eq:amat2kp}
\end{equation}
where the matrix $R_n$ is given by (\ref{subeqs:rkp}). 
If we put $t_n\equiv 0\ (n\ge r)$, then we have 
$R_k\equiv 0\ (k\ge r-1)$, 
and $A$ has a pole of degree $r$ at $\lambda=\infty$. 
In this case, 
the linear system (\ref{eq:linakp}) is said to have an irregular 
singular point at $\lambda=\infty$ of Poincar\'e rank $r-1$. 

By using (\ref{eq:bmatkp}) and (\ref{eq:amat2kp}), 
the left-hand side of the system (\ref{eq:cckp}) becomes
\begin{equation}
\begin{split}
&\partial_{a_n} A-\partial_{\lambda}C_n+[A,C_n]\\
&=\left(\partial_{a_n}P+\left[\frac{P}{a_n},R_n\right]\right)\frac{1}{\lambda}
+\left(\partial_{a_n}Q+\left[\frac{Q}{a_n-1},R_n\right]\right)
\frac{1}{\lambda-1}\\
&\quad{}+\sum_{
\begin{subarray}{c}
m=1,\dots,l\\
m\ne n
\end{subarray}}
\left(\partial_{a_n}R_m+\left[\frac{R_m}{a_n-a_m},R_n\right]\right)
\frac{1}{\lambda-a_m}\\
&\quad{}+\left(\partial_{a_n}R_n-\left[\frac{P}{a_n}+\frac{Q}{a_n-1}
+\sum_{
\begin{subarray}{c}
m=1,\dots,l\\
m\ne n
\end{subarray}}
\frac{R_m}{a_n-a_m}
+\sum_{l=0}^{\infty}{a_n}^lT_l,R_n\right]\right)\frac{1}{\lambda-a_n}\\
&\quad{}+\sum_{k=0}^{\infty}\left(\partial_{a_n}T_k
-\left[\sum_{l=k+1}^{\infty}{a_n}^{l-k-1}T_l,R_n\right]
\right)\lambda^k. 
\end{split}
\end{equation}
It follows that we obtain the systems
\begin{subequations}\label{subeq:p6exkp}
\begin{align}
\partial_{a_n}P+\left[\frac{P}{a_n},R_n\right]&=0,\\
\partial_{a_n}Q+\left[\frac{Q}{a_n-1},R_n\right]&=0,\\
\partial_{a_n}R_m+\left[\frac{R_m}{a_n-a_m},R_n\right]&=0\: (m\ne n),\\
\partial_{a_n}R_n-\left[\frac{P}{a_n}+\frac{Q}{a_n-1}
+\sum_{
\begin{subarray}{c}
m=1,\dots,l\\
m\ne n
\end{subarray}}
\frac{R_m}{a_n-a_m}+\sum_{l=0}^{\infty}{a_n}^lT_l,R_n\right]&=0,\\
\partial_{a_n}T_k
-\left[\sum_{l=k+1}^{\infty}{a_n}^{l-k-1}T_l,R_n\right]
&=0\ (k\ge 0).
\end{align}
\end{subequations}
If we put $t_n\equiv 0\  (n\ge 1)$ and $x\equiv 0$, 
then the coefficient matrices reduce to 
$T_k\equiv 0\  (k\ge 0)$ 
and we have 
\begin{subequations}\label{subeq:sch}
\begin{align}
\partial_{a_n}P+\left[\frac{P}{a_n},R_n\right]&=0,\\
\partial_{a_n}Q+\left[\frac{Q}{a_n-1},R_n\right]&=0,\\
\partial_{a_n}R_m+\left[\frac{R_m}{a_n-a_m},R_n\right]&=0\quad (m\ne n),\\
\partial_{a_n}R_n-\left[\frac{P}{a_n}+\frac{Q}{a_n-1}
+\sum_{
\begin{subarray}{c}
m=1,\dots,l\\
m\ne n
\end{subarray}}
\frac{R_m}{a_n-a_m},R_n\right]&=0.
\end{align}
\end{subequations}
This system is nothing but the Schlesinger system (\cite{S}).
If we set $l=1$, then we have
\begin{subequations}\label{subeqs:p6kp}
\begin{align}
\partial_{a_1}P+\left[\frac{P}{a_1},R_1\right]&=0,\\
\partial_{a_1}Q+\left[\frac{Q}{a_1-1},R_1\right]&=0.
\end{align}
\end{subequations}
This system is equivalent to $\mathrm{P_{VI}}$ in the paper, \cite{JM}.

\section{The two-component KP hierarchy and the other Painlev\'e 
equations}
\label{sec:otherkp}
In this section, 
we study holonomic deformation 
relating to the $(1,1)$-re\-duc\-tion of 
the two-component KP hierarchy. 
We show that systems obtained from the deformation reduces to the 
Painlev\'e equation, 
$\mathrm{P_{V}}$, $\mathrm{P_{IV}}$, $\mathrm{P_{III}}$ and 
$\mathrm{P_{II}}$.

\subsection{The fifth Painlev\'e equation}\label{subsec:fifth}
We explain the $(1,1)$-reduction of the two-component KP hierarchy.
We show that the systems that describes the condition of 
the holonomic deformation that contains this hierarchy 
as a part reduces to $\mathrm{P_{V}}$.
Therefore we find that $\mathrm{P_{V}}$ is obtained 
through the reduction from the nonlinear Schr\"odinger equation.

We define the gauge operator
\begin{equation}
\mathcal{W}
=I+\sum_{k=1}^{\infty}w_{k}\partial_x^{-k}
\end{equation}
whose $2\times 2$ coefficients matrices $w_{k}$ do not depend on 
the parameter $x$.
This condition for the coefficients is equivalent to 
\lq\lq the $(1,1)$-reduction".
By using the gauge operator $\mathcal{W}$, 
we define a pseudo-differential operator $\mathcal{U}$ by
\begin{align}
\begin{split}
\mathcal{U}&=\mathcal{W}\sigma_3\mathcal{W}^{-1}
=\sigma_3+\sum_{k=1}^{\infty}u_k\partial_x^{-k}.\label{eq:m5kp}
\end{split}
\end{align}
We define a differential operator $\mathcal{B}_n$ by 
\begin{equation}
\mathcal{B}_n
=\left(\mathcal{W}\sigma_3\partial_x^n\mathcal{W}^{-1}\right)_+
=\sum_{k=0}^{n-1}u_{n-k}\partial_x^{k}+\sigma_3\partial_x^{n}
\quad(n\ge 1).
\end{equation}
Matrix operators
\begin{align}
W&=I+\sum_{k=1}^{\infty}w_{k}\lambda^{-k},\\
U&=\sigma_3+\sum_{k=1}^{\infty}u_k\lambda^{-k},\label{eq:umatkp5}\\
B_n&=\sum_{k=0}^{n-1}u_{n-k}\lambda^{k}
+\sigma_3\lambda^{n}\quad(n\ge 1)\label{eq:bmatkp5}
\end{align}
are obtained from the pseudo-differential operators by replacing 
$\partial_x$ with $\lambda$.
We assume that the matrix operators satisfy the Sato equation
\begin{equation}
\partial_{t_n} W=B_nW-W\sigma_3\lambda^n\quad(n\ge 1).
\label{eq:sato5kp}
\end{equation}

We define a wave function
\begin{equation}
\Psi(\lambda)=W\Psi_0(\lambda),\label{eq:wave5kp}
\end{equation}
where
\begin{equation}
\begin{split}
\Psi_0(\lambda)
&=\lambda^{\alpha}(\lambda-1)^{\beta}\exp(x\lambda)\\
&\quad{}\times
\begin{pmatrix}
\lambda^{a}(\lambda-1)^{b}\exp(\sum_{n=1}^{\infty}t_n\lambda^n)&0\\
0&\lambda^{-a}(\lambda-1)^{-b}\exp(-\sum_{n=1}^{\infty}t_n\lambda^n)
\end{pmatrix}.
\end{split}\label{eq:psi05kp}
\end{equation}
This definition of the wave function is slightly different from 
the usual one. 
The element of $\Psi_0(\lambda)$ is similar to the integrand of 
the integral representation of 
the confluent hypergeometric function: 
\begin{equation}
{}_1F_1(a;b;t)
=\frac{\Gamma(b)}{\Gamma(b-a)\Gamma(a)}
\int_0^1\lambda^{a-1}(1-\lambda)^{b-a-1}
e^{t \lambda}d\lambda.
\end{equation}
The difference does not affect the soliton system, but affects 
the system of the holonomic deformation.
We note that the matrix-valued function $\Psi_0(\lambda)$ satisfies 
\begin{align}
\partial_x\Psi_0(\lambda)&=\lambda\Psi_0(\lambda),\\
\partial_{t_n}\Psi_0(\lambda)
&=\sigma_3 \lambda^n\Psi_0(\lambda)
=\sigma_3\partial_x^n\Psi_0(\lambda)\quad(n\ge 1).
\end{align}

This leads to the following proposition: 
\begin{prop}\label{prop:lax5kp}
If a matrix operator $W$ satisfies 
the Sato equation $(\ref{eq:sato5kp})$, 
then the matrix operators $U$ and $B_n$ 
satisfy 
\begin{gather}
\partial_{t_n} U=[B_n,U]\quad (n\ge 1),\\
\partial_{t_m}B_n-\partial_{t_n}B_m+[B_n,B_m]=0\quad (n,m\ge 1).
\label{eq:zs2kp5}
\end{gather}
Furthermore, 
the wave function $\Psi(\lambda)$ satisfies the linear systems, 
\begin{align}
\partial_x\Psi(\lambda)
&=\lambda\Psi(\lambda),\label{eq:linl5kp} \\ 
\partial_{t_n} \Psi(\lambda)
&=B_n\Psi(\lambda)\quad(n\ge 1).\label{eq:linb5kp}
\end{align}
\end{prop}

If we choose $m=1$ and $n=2$, then the Zakharov-Shabat system 
(\ref{eq:zs2kp5})
\begin{equation}
\partial_{t_1}B_2-\partial_{t_2}B_1+[B_2,B_1]=0
\end{equation}
yields
\begin{subequations}
\begin{align}
\partial_{t_1} u_1+[u_2,\sigma_3]&=0,\\
\partial_{t_1} u_2-\partial_{t_2}u_1+[u_2,u_1]&=0.
\end{align}
\end{subequations}
If we use the following parameterizations for the matrices 
\begin{subequations}
\begin{align}
u_1&=\begin{pmatrix}
0&u\\
v&0
\end{pmatrix},\\
u_2&=\begin{pmatrix}
-uv/2&f\\
g&uv/2
\end{pmatrix},
\end{align}
\end{subequations}
then we have the nonlinear Schr\"o\-dinger equation 
\begin{subequations}\label{eqs:solkp5}
\begin{align}
\partial_{t_1}u-2f&=0,\\
\partial_{t_1}v+2g&=0,\\
\partial_{t_1}f-\partial_{t_2}u-u^2v&=0,\\
\partial_{t_1}g-\partial_{t_2}v+uv^2&=0.
\end{align}
\end{subequations}

\bigskip

We consider the holonomic deformation that contains 
the two-component system. 
If we introduce a differential operator
\begin{equation}
\begin{split}
\mathcal{V}
&=\alpha-\beta\sum_{k=1}^\infty\partial_x^k
+x\partial_x
+\sigma_3\left\{a-b\sum_{k=1}^\infty\partial_x^k
+\sum_{n=1}^{\infty}nt_n\partial_x^n\right\},
\end{split}
\end{equation}
then the matrix-valued function $\Psi_0(\lambda)$ 
(\ref{eq:psi05kp}) satisfies 
\begin{equation}
\lambda\partial_{\lambda}\Psi_0(\lambda)=\mathcal{V}\Psi_0(\lambda).
\end{equation}
By using the gauge operator $\mathcal{W}$ 
and the differential operator $\mathcal{V}$,
we define a differential operator $\mathcal{D}$ by 
\begin{equation}
\mathcal{D}=
\left(\mathcal{W}\mathcal{V}\mathcal{W}^{-1}\right)_{+}
=\sum_{k=0}^{\infty}d_ke^{k\partial_s},
\end{equation}
where
\begin{subequations}
\begin{align}
d_0&=\alpha I+a\sigma_3-b\sum_{l=1}^{\infty} u_l
+\sum_{n=1}^{\infty}nt_nu_n,\\
d_1&=(-\beta+x)I
-b\left(\sigma_3+\sum_{l=1}^{\infty} u_l\right)
+t_1\sigma_3+\sum_{n=2}^{\infty}nt_nu_{n-1},\\
d_k&=-\beta I-b\left(\sigma_3+\sum_{l=1}^{\infty} u_l\right)
+kt_{k}\sigma_3
+\sum_{n=k+1}^{\infty}nt_nu_{n-k}\quad(k\ge 2).
\end{align}
\end{subequations}
We introduce matrix operators
\begin{gather}
T=\frac{\alpha I+a\sigma_3}{\lambda}
+\frac{\beta I+b\sigma_3}{\lambda-1}
+\sum_{n=1}^{\infty}nt_n\sigma_3\lambda^{n-1},\\ 
A=
\sum_{k=0}^{\infty}d_k\lambda^{k-1}.\label{eq:amatkp5}
\end{gather}
We note that 
\begin{equation}
\partial_{\lambda}\Psi_0(\lambda)=T\Psi_0(\lambda).
\end{equation}
We assume that the matrix operator $A$ satisfies the condition 
\begin{equation}\label{eq:satoa5kp}
\partial_\lambda W=AW-WT.
\end{equation}

This leads to the following proposition: 
\begin{prop}\label{prop:laxa5kp}
If a matrix operator $W$ satisfies the reduction condition 
$(\ref{eq:satoa5kp})$, 
then the matrix operators $U$, $A$ and $B_n$ 
satisfy 
\begin{gather}
\partial_{\lambda} U=[A,U],\\
\partial_{t_n}A-\partial_{\lambda}B_n+[A,B_n]=0\quad (n\ge 1).
\label{eq:cc5kp}
\end{gather}
Furthermore, 
the wave function $\Psi(\lambda)$ $(\ref{eq:wave5kp})$ 
satisfies the linear system,
\begin{equation}\label{eq:lina5kp}
\partial_\lambda \Psi(\lambda)=A\Psi(\lambda).
\end{equation}
\end{prop}

If we introduce matrices
\begin{subequations}
\begin{align}
\begin{split}
P&=\alpha I+a\sigma_3-b\sum_{l=1}^{\infty} u_l
+\sum_{n=1}^{\infty}nt_nu_n,
\end{split}\\
\begin{split}
Q&=\beta I+b\left(\sigma_3+\sum_{l=1}^{\infty} u_l\right),
\end{split}\\
\begin{split}
T_0&=xI+t_1\sigma_3+\sum_{n=2}nt_n u_{n-1},
\end{split}\\
\begin{split}
T_k&=(k+1)t_{k+1}\sigma_3+\sum_{n=k+2}nt_n u_{n-k-1}\quad(k\ge 1), 
\end{split}
\end{align}
\end{subequations}
then we have 
\begin{equation}
\begin{split}
A&=\frac{P}{\lambda}+\frac{Q}{\lambda-1}
+\sum_{k=0}^{\infty}T_k\lambda^k.
\end{split}\label{eq:amat2kp5}
\end{equation}
By using (\ref{eq:bmatkp5}) and (\ref{eq:amat2kp5}), 
the left-hand side of the system (\ref{eq:cc5kp}) with $n=1$ 
turns
\begin{equation}
\begin{split}
\partial_{t_1} A-\partial_{\lambda}B_1&+[A,B_1]\\
&=\left(\partial_{t_1} P+[P,u_1]\right)\frac{1}{\lambda}
+\left(\partial_{t_1} Q+[Q,u_1+\sigma_3]\right)\frac{1}{\lambda-1}\\
&\quad{}+\partial_{t_1} T_0-\sigma_3+[T_0,u_1]+[P+Q,\sigma_3]\\
&\quad{}
+\sum_{k=1}^{\infty}\left(\partial_{t_1} T_k+[T_k,u_1]
+[T_{k-1},\sigma_3]\right)\lambda^k. 
\end{split}
\end{equation}
Therefore we obtain the systems
\begin{subequations}\label{subeq:p5exkp}
\begin{align}
\partial_{t_1} P+[P,u_1]&=0,\\
\partial_{t_1} Q+[Q,u_1+\sigma_3]&=0,\\
\partial_{t_1} T_0-\sigma_3+[T_0,u_1]+[P+Q,\sigma_3]&=0,\\
\partial_{t_1} T_k+[T_k,u_1]+[T_{k-1},\sigma_3]&=0\quad(k\ge 1).
\end{align}
\end{subequations}
If we put $t_n\equiv 0\, (n\ge 2)$, 
then the coefficient matrices reduce to
$T_0=t_1\sigma_3$, $T_k\equiv 0\ (k\ge 1)$, 
and then we have 
\begin{subequations}
\begin{align}
\partial_{t_1} P+[P,u_1]&=0,\\
\partial_{t_1} Q+[Q,u_1+\sigma_3]&=0.
\end{align}
\end{subequations}
This systems is equivalent to $\mathrm{P_{V}}$ in the paper, 
\cite{JM}.

\begin{rem}
We can also formulate this hierarchy by using the difference 
operators (\cite{UT}). 
If the gauge operator $\mathcal{W}$ do not depend on the 
parameter $\alpha$, then we have
\begin{equation}
e^{\partial_{\alpha}}\Psi(\lambda)=\lambda\Psi(\lambda).
\end{equation}
So the difference operators are obtained from 
the pseudo-differential operators by replacing 
$\partial_x$ with $e^{\partial_{\alpha}}$.
\end{rem}

\subsection{The fourth Painlev\'e equation}
We consider the different holonomic deformation that relates to 
the hierarchy in the previous subsection. 
We show that the system that describes the deformation condition
reduces to $\mathrm{P_{IV}}$.
This fact follows the result in the paper, \cite{JM2}.

We employ the same soliton system in the previous subsection. 
But we define the wave function as follows:
\begin{equation}
\Psi(\lambda)=W\Psi_0(\lambda),\label{eq:wave4kp}
\end{equation}
where
\begin{equation}
\Psi_0(\lambda)
=\lambda^{\alpha}\exp(x\lambda)
\begin{pmatrix}
\lambda^{a}\exp\left(\sum_{n=1}^{\infty}t_n\lambda^n\right)&0\\
0&\lambda^{-a}\exp\left(-\sum_{n=1}^{\infty}t_n\lambda^n\right)
\end{pmatrix}.\label{eq:psi04kp}
\end{equation}
The element of $\Psi_0(\lambda)$ is similar to the integrand of 
the integral representation of 
the Hermite-Weber function: 
\begin{equation}
H_{\nu}(t)
=\frac{\Gamma(\nu+1)}{2\pi i}
\int_C \lambda^{-\nu-1}e^{2t\lambda-\lambda^2}d\lambda.
\end{equation}
The matrix-valued function $\Psi_0(\lambda)$ satisfies 
\begin{align}
\partial_x\Psi_0(\lambda)&=\lambda\Psi_0(\lambda),\\
\partial_{t_n}\Psi_0(\lambda)
&=\sigma_3 \lambda^n\Psi_0(\lambda)
=\sigma_3\partial_x^n\Psi_0(\lambda)\quad(n\ge 1).
\end{align}

This leads to the following proposition: 
\begin{prop}\label{prop:lax4kp}
If a matrix operator $W$ satisfies 
the Sato equation $(\ref{eq:sato5kp})$,  
then the wave function $\Psi(\lambda)$ satisfies the linear systems, 
\begin{align}
\partial_x\Psi(\lambda)
&=\lambda\Psi(\lambda),\label{eq:linl4kp} \\ 
\partial_{t_n} \Psi(\lambda)
&=B_n\Psi(\lambda)\quad(n \ge 1).\label{eq:linb4kp}
\end{align}
\end{prop}

\bigskip

We present the reduction condition for the soliton system.
If we introduce a differential operator
\begin{equation}
\mathcal{T}
=I(\alpha+x\partial_x)
+\sigma_3\left(a+\sum_{n=1}^{\infty}nt_n\partial_x^{n}\right),
\end{equation}
then the matrix-valued function $\Psi_0(\lambda)$ 
(\ref{eq:psi04kp}) satisfies 
\begin{equation}
\lambda\partial_{\lambda}\Psi_0(\lambda)=\mathcal{T}\Psi_0(\lambda).
\end{equation}
By using the gauge operator $\mathcal{W}$ 
and the differential operator $\mathcal{T}$,
we define a differential operator $\mathcal{A}$ by 
\begin{equation}
\mathcal{A}=
\left(\mathcal{W}\mathcal{T}\mathcal{W}^{-1}\right)_{+}
=\sum_{k=0}^{\infty}a_k\partial_x^{k},
\end{equation}
where
\begin{subequations}
\begin{align}
a_0&=\alpha I+a\sigma_3+\sum_{n=1}^{\infty}nt_n u_n,\\
a_1&=xI+t_1\sigma_3+\sum_{n=2}^{\infty}nt_n u_{n-1},\\
a_k&=kt_k\sigma_3+\sum_{n=k+1}^{\infty}nt_n u_{n-k}\quad(k\ge 2).
\end{align}
\end{subequations}
We introduce matrix operators
\begin{gather}
T=I(\alpha+x\lambda) 
+\sigma_3\left(a+\sum_{n=1}^{\infty}nt_n\lambda^{n}\right),\\
A=\sum_{k=0}^{\infty}a_k\lambda^{k}\label{eq:amatkp4}.
\end{gather}
We assume that the matrix operator $A$ satisfies 
\begin{equation}\label{eq:satoa4kp}
\lambda\partial_\lambda W=AW-WT.
\end{equation}

This leads to the following proposition: 
\begin{prop}\label{prop:laxa4kp}
If a matrix operator $W$ satisfies the reduction condition 
$(\ref{eq:satoa4kp})$, 
then the matrix operators $U$, $A$ and $B_n$
satisfy 
\begin{gather}
\lambda\partial_{\lambda} U=[A,U],\\
\partial_{t_n}A-\lambda\partial_{\lambda}B_n+[A,B_n]=0\quad (n\ge 1).
\label{eq:laxkp4}
\end{gather}
Furthermore, the wave function $\Psi(\lambda)$ $(\ref{eq:wave4kp})$ 
satisfies the linear system,
\begin{equation}\label{eq:lina4kp}
\lambda\partial_\lambda \Psi(\lambda)=A\Psi(\lambda).
\end{equation}
\end{prop}
\begin{rem}
If we put $t_n \equiv 0\ (n\ge l)$, 
then we have $a_k\equiv 0\ (k\ge l)$.  
In this case, the linear system (\ref{eq:lina4kp}) has a regular 
singular point at $\lambda=0$ and an irregular singular point 
at $\lambda=\infty$ of Poincar\'e rank $l-1$. 
Hence we guess that the systems (\ref{eq:laxkp4}) are 
equivalent to the fourth Painlev\'e equation with several variables; 
see \cite{K1,K2,K3}.
\end{rem}
By using (\ref{eq:bmatkp5}) and (\ref{eq:amatkp4}), 
the left-hand side of the system (\ref{eq:laxkp4}) with $n=1$ turns
\begin{equation}
\begin{split}
\partial_{t_1}A&-\lambda\partial_{\lambda}B_1+[A,B_1]\\
&=\partial_{t_1} a_0+\left[a_0,u_1\right]+
\left(\partial_{t_1}a_1-\sigma_3+[a_1,u_1]+[a_0,\sigma_3]\right)
\lambda\\
&\quad{}+\sum_{k=2}^{\infty}
\left(\partial_{t_1} a_k+[a_k,u_1]+[a_{k-1},\sigma_3]\right)\lambda^k. 
\end{split}
\end{equation}
Hence we have the systems
\begin{subequations}\label{subeq:p4exkp}
\begin{align}
\partial_{t_1} a_0+\left[a_0,u_1\right]&=0,\\
\partial_{t_1} a_1-\sigma_3+[a_1,u_1]+[a_0,\sigma_3]&=0,\\
\partial_{t_1} a_k+[a_k,u_1]+[a_{k-1},\sigma_3]&=0\quad(k\ge 2).
\end{align}
\end{subequations}
If we put $t_2\equiv 1/2$, $t_n \equiv 0\ (n\ge 3)$, 
then the coefficient matrices reduce to
$a_2=\sigma_3$, $a_k\equiv 0\ (k\ge 3)$, 
and we have 
\begin{subequations}
\begin{align}
\partial_{t_1} a_0+\left[a_0,u_1\right]&=0,\\
\partial_{t_1} a_1-\sigma_3+[a_1,u_1]+\left[a_0,\sigma_3\right]&=0.
\end{align}
\end{subequations}
This systems is equivalent to $\mathrm{P_{IV}}$ in the paper, 
\cite{JM}.

\subsection{The third Painlev\'e equation}
We present that the system that is the condition of 
the different holonomic deformation 
reduces to $\mathrm{P_{III}}$.
So we find that $\mathrm{P_{III}}$ is obtained 
through the reduction from the nonlinear Schr\"odinger equation.

We employ the same soliton system in the previous subsection, 
and we give another reduction condition for the soliton system.
If we introduce a differential operator 
\begin{equation}
\mathcal{T}
=I\left(\alpha\partial_x+x\partial_x^2\right)
+\sigma_3\left(a \partial_x
+\sum_{n=1}^{\infty}nt_n\partial_x^{n+1}\right),
\end{equation}
then the matrix-valued function $\Psi_0(\lambda)$ 
(\ref{eq:psi04kp}) satisfies 
\begin{equation}
\lambda^2\partial_{\lambda}\Psi_0(\lambda)
=\mathcal{T}\Psi_0(\lambda).
\end{equation}
By using the gauge operator $\mathcal{W}$ 
and the differential operator $\mathcal{T}$,
we define a differential operator $\mathcal{A}$ by 
\begin{equation}
\mathcal{A}=
\left(\mathcal{W}\mathcal{T}\mathcal{W}^{-1}\right)_{+}
=\sum_{k=0}^{\infty}a_k\partial_x^{k},
\end{equation}
where
\begin{subequations}
\begin{align}
a_0&=-w_1+a u_1+\sum_{n=1}^{\infty}nt_n u_{n+1},\\
a_1&=\alpha I+a \sigma_3+\sum_{n=1}^{\infty}nt_n u_{n},\\
a_2&=xI+t_1\sigma_3+\sum_{n=2}^{\infty}nt_n u_{n-1},\\
a_k&=(k-1)t_{k-1}\sigma_3+\sum_{n=k}^{\infty}nt_n u_{n-k+1}
\quad(k\ge 3).
\end{align}
\end{subequations}
We introduce matrix operators
\begin{gather}
T=I\left(\alpha \lambda+x\lambda^2\right)
+\sigma_3\left(a \lambda
+\sum_{n=1}^{\infty}nt_n\lambda^{n+1}\right),\\
A=\sum_{k=0}^{\infty}a_k\lambda^{k}.\label{eq:amatkp3}
\end{gather}
We assume that the matrix operator $A$ satisfies 
\begin{equation}\label{eq:satoa3kp}
\lambda^2\partial_\lambda W=AW-WT.
\end{equation}

This leads to the following proposition: 
\begin{prop}\label{prop:laxa3kp}
If a matrix operator $W$ satisfies the reduction condition 
$(\ref{eq:satoa3kp})$, 
then the matrix operators $U$, $A$ and $B_n$
satisfy 
\begin{gather}
\lambda^2\partial_{\lambda} U=[A,U],\\
\partial_{t_n}A-\lambda^2\partial_{\lambda}B_n+[A,B_n]=0
\quad(n\ge 1).
\label{eq:laxkp3}
\end{gather}
Furthermore, the wave function $\Psi(\lambda)$ $(\ref{eq:wave4kp})$ 
satisfies the linear system,
\begin{equation}\label{eq:lina3kp}
\lambda^2\partial_\lambda \Psi(\lambda)=A\Psi(\lambda).
\end{equation}
\end{prop}

By using (\ref{eq:bmatkp5}) and (\ref{eq:amatkp3}), 
the left-hand side of the system (\ref{eq:laxkp3}) with $n=1$ is
\begin{equation}
\begin{split}
\partial_{t_1} A&-\lambda^2\partial_{\lambda}B_1+[A,B_1]\\
&=\partial_{t_1} a_0+[a_0,u_1]
+\left(\partial_{t_1} a_1+[a_1,u_1]+[a_0,\sigma_3]\right)\lambda\\
&\quad {}
+(\partial_{t_1}a_2-\sigma_3+[a_2,u_1]+[a_1,\sigma_3])\lambda^2\\
&\quad {}
+\sum_{k=3}^{\infty}(\partial_{t_1}a_k+[a_k,u_1]+[a_{k-1},\sigma_3])
\lambda^k.
\end{split}
\end{equation}
Thus we obtain the systems
\begin{subequations}\label{subeqs:p3exkp}
\begin{align}
\partial_{t_1} a_0+[a_0,u_1]&=0,\\
\partial_{t_1} a_1+[a_1,u_1]+[a_0,\sigma_3]&=0,\\
\partial_{t_1} a_2-\sigma_3+[a_2,u_1]+[a_1,\sigma_3]&=0,\\
\partial_{t_1} a_k+[a_k,u_1]+[a_{k-1},\sigma_3]&=0\quad(k\ge 3).
\end{align}
\end{subequations}
If we put $t_n\equiv 0\ (n\ge 2)$, 
then the coefficient matrices reduce to
$a_2=t_1\sigma_3,\,a_k\equiv 0\ (k\ge 3)$, 
and then we have 
\begin{subequations}\label{eqs:pmatkp3}
\begin{align}
\partial_{t_1} a_0+[a_0,u_1]&=0,\\
\partial_{t_1} a_1+[a_1,u_1]+[a_0,\sigma_3]&=0.
\end{align}
\end{subequations}
We can obtain $\mathrm{P_{III}}$ from this system (\ref{eqs:pmatkp3}).

\subsection{The second Painlev\'e equation}
We present that the system that describes the condition of 
the different holonomic deformation 
reduces to $\mathrm{P_{II}}$.

We employ the same soliton system in Subsection~\ref{subsec:fifth}. 
However we define the wave function as follows:
\begin{equation}
\Psi(\lambda)=W\Psi_0(\lambda),\label{eq:wave2kp}
\end{equation}
where
\begin{equation}
\Psi_0(\lambda)
=\lambda^\alpha e^{x\lambda}
\begin{pmatrix}
\exp\left(\sum_{n=1}^{\infty}t_n\lambda^n\right)&0\\
0&\exp\left(-\sum_{n=1}^{\infty}t_n\lambda^n\right)
\end{pmatrix}.\label{eq:psi02kp}
\end{equation}
Needless to say, 
the matrix-valued function $\Psi_0(\lambda)$ satisfies 
\begin{align}
\partial_x\Psi_0(\lambda)&=\lambda\Psi_0(\lambda),\\
\partial_{t_n}\Psi_0(\lambda)
&=\sigma_3 \lambda^n\Psi_0(\lambda)
=\sigma_3\partial_x^n\Psi_0(\lambda)\quad(n\ge 1).
\end{align}

This leads to the following proposition: 
\begin{prop}\label{prop:lax2kp}
If a matrix operator $W$ satisfies 
the Sato equation $(\ref{eq:sato5kp})$,  
then the wave function $\Psi(\lambda)$ satisfies the linear systems, 
\begin{align}
\partial_x\Psi(\lambda)&=\lambda\Psi(\lambda),\label{eq:linl2kp} \\ 
\partial_{t_n} \Psi(\lambda)
&=B_n\Psi(\lambda)\quad(n \ge 1).\label{eq:linb2kp}
\end{align}
\end{prop}

\bigskip

We give the reduction condition for the soliton system.
If we introduce a differential operator
\begin{equation}
\mathcal{T}=I\left(\alpha\partial_x^{-1}+x\right)
+\sigma_3\sum_{n=1}^{\infty}nt_n\partial_x^{n-1},
\end{equation}
then the matrix-valued function $\Psi_0(\lambda)$ 
(\ref{eq:psi02kp}) satisfies 
\begin{equation}
\partial_{\lambda}\Psi_0(\lambda)=\mathcal{T}\Psi_0(\lambda).
\end{equation}
By using the gauge operator $\mathcal{W}$ 
and the differential operator $\mathcal{T}$,
we define a differential operator $\mathcal{A}$ by 
\begin{equation}
\mathcal{A}=
\left(\mathcal{W}\mathcal{T}\mathcal{W}^{-1}\right)_{+}
=\sum_{k=0}^{\infty}a_ke^{k\partial_s},
\end{equation}
where
\begin{align}
a_0&=xI+t_1\sigma_3+\sum_{n=2}^{\infty}nt_n u_{n-1},\\
a_k&=(k+1)t_{k+1}\sigma_3+\sum_{n=k+2}^{\infty}nt_n u_{n-k-1}
\quad(k\ge 1).
\end{align}
We introduce matrix operators
\begin{gather}
T=I(\alpha \lambda^{-1}+x)
+\sigma_3\sum_{n=1}^{\infty}nt_n\lambda^{n-1},\\
A=\sum_{k=0}^{\infty}a_k\lambda^{k}\label{eq:amatkp2}.
\end{gather}
We assume that the matrix operator $A$ satisfies 
\begin{equation}\label{eq:satoa2kp}
\partial_\lambda W=AW-WT.
\end{equation}

This leads to the following proposition: 
\begin{prop}\label{prop:laxa2kp}
If a matrix operator $W$ satisfies the reduction condition 
$(\ref{eq:satoa2kp})$, 
then the matrix operators $U$, $A$ and $B_n$
satisfy 
\begin{gather}
\partial_{\lambda} U=[A,U],\\
\partial_{t_n}A-\partial_{\lambda}B_n+[A,B_n]=0\quad (n\ge 1).
\label{eq:laxkp2}
\end{gather}
Furthermore, the wave function $\Psi(\lambda)$ $(\ref{eq:wave2kp})$ 
satisfies the linear system,
\begin{equation}\label{eq:lina2kp}
\partial_\lambda \Psi(\lambda)=A\Psi(\lambda).
\end{equation}
\end{prop}
\begin{rem}
If we put $t_n \equiv 0\ (n\ge l)$, 
then we have $a_k\equiv 0\ (k\ge l-1)$.  
In this case, the linear system (\ref{eq:lina2kp}) has 
an irregular singular point 
at $\lambda=\infty$ of Poincar\'e rank $l-1$. 
So we guess that the systems (\ref{eq:laxkp2}) are 
equivalent to the $A_g$-system; 
see \cite{L1,L2,L3}.
\end{rem}
By using (\ref{eq:bmatkp5}) and (\ref{eq:amatkp2}), 
the left-hand side of the system (\ref{eq:laxkp2}) with $n=1$ turns
\begin{equation}
\begin{split}
\partial_{t_1} A&-\partial_{\lambda}B_1+[A,B_1]\\
&=\partial_{t_1}a_0-\sigma_3+[a_0,u_1]
+\sum_{k=1}^{\infty}
\left(\partial_{t_1}a_k+[a_k,u_1]+[a_{k-1},\sigma_3]\right)\lambda^k. 
\end{split}
\end{equation}
So we have the systems
\begin{subequations}\label{subeqs:p2exkp}
\begin{align}
\partial_{t_1}a_0-\sigma_3+[a_0,u_1]&=0,\\
\partial_{t_1}a_k+[a_k,u_1]+[a_{k-1},\sigma_3]&=0\quad(k\ge 1).
\end{align}
\end{subequations}
If we put $t_3\equiv 1/3$, $t_n\equiv 0\ (n=2,\,n\ge 4)$, 
then the coefficient matrices reduce to
$a_2=\sigma_3$, $a_k\equiv 0\ (k\ge 3)$, 
and we have 
\begin{subequations}
\begin{align}
\partial_{t_1}a_0-\sigma_3+[a_0,u_1]&=0,\\
\partial_{t_1}a_1+[a_1,u_1]+[a_0,\sigma_3]&=0.
\end{align}
\end{subequations}
This systems is equivalent to $\mathrm{P_{II}}$ in the paper, 
\cite{JM}.

\subsection*{Acknowledgments}
The author wishes to thank 
Professor Michio Jimbo, 
Professor Saburo Kakei, 
Professor Masatoshi Noumi, 
Professor Yousuke Ohyama, 
Professor Kazuo Okamoto, 
Professor Hidetaka Sakai, 
Professor Tetsuji Tokihiro, 
Doctor Teru\-hisa Tsuda and 
Professor Ralph Willox 
for valuable comments. 
This work is partially supported by the Japan Society for 
the Promotion of Science (JSPS). 

\bibliographystyle{plain}

\end{document}